\documentclass[%
 reprint,
superscriptaddress,
 amsmath,amssymb,
 aps,
prb,
]{revtex4-2}
\usepackage{graphicx}
\usepackage{dcolumn}
\usepackage{bm}
\usepackage[final]{changes}

\usepackage{pdfpages}
\usepackage{pgffor}
\makeatletter
\AtBeginDocument{\let\LS@rot\@undefined}
\makeatother

\begin{document}

\preprint{APS/123-QED}

\title{Gain/Loss-free Non-Hermitian Metamaterials}

\author{Maopeng Wu}
\thanks{These two authors contributed equally.}
	\affiliation{State Key Laboratory of Tribology in Advanced Equipment,
 Department of Mechanical Engineering,
Tsinghua University, Beijing 100084, China}	

\author{Mingze Weng}
\thanks{These two authors contributed equally.}
	\affiliation{State Key Laboratory of Tribology in Advanced Equipment, 
Department of Mechanical Engineering,
Tsinghua University, Beijing 100084, China}	

\author{Zhonghai Chi}
	\affiliation{State Key Laboratory of New Ceramics and Fine Processing,\\
School of Materials Science and Engineering, Tsinghua University, Beijing 100084, China}

\author{Siyong Zheng}
	\affiliation{State Key Laboratory of New Ceramics and Fine Processing,\\
School of Materials Science and Engineering, Tsinghua University, Beijing 100084, China}
\author{Fubei Liu}
	\affiliation{State Key Laboratory of New Ceramics and Fine Processing,\\
School of Materials Science and Engineering, Tsinghua University, Beijing 100084, China}

\author{Weijia Luo}
	\affiliation{State Key Laboratory of New Ceramics and Fine Processing,\\
School of Materials Science and Engineering, Tsinghua University, Beijing 100084, China}

\author{Qian Zhao}%
 	\email{zhaoqian@tsinghua.edu.cn}
 	\affiliation{State Key Laboratory of Tribology in Advanced Equipment, 
Department of Mechanical Engineering,
Tsinghua University, Beijing 100084, China}

\author{Yonggang Meng}%
 	\affiliation{State Key Laboratory of Tribology in Advanced Equipment, 
Department of Mechanical Engineering,
Tsinghua University, Beijing 100084, China}
\author{Ji Zhou}
	\email{zhouji@tsinghua.edu.cn}
	\affiliation{State Key Laboratory of New Ceramics and Fine Processing,\\
School of Materials Science and Engineering, Tsinghua University, Beijing 100084, China}
\date{\today}

\begin{abstract}
The ease of using optical gain/loss provides a fertile ground for experimental explorations of non-Hermitian (NH) physics. Without gain/loss, can we realize the NH effect in a Hermitian system? The interface between the coupled Hermitian subsystems is a natural object for NH physics due to the nonconservative process on it. However, it is still far from enduing the interface with rich NH physics. Here, a junction between the topological insulator and the conductor is considered, where the interface can be effectively described by a NH Hamiltonian\textemdash such NH character is ascribed to the conductor self-energy of a reservoir. As a consequence of that, we show the wave propagation along the interface exhibits dissipative non-reciprocity (dubbed non-Bloch transport), which was believed to be unique in NH systems. Moreover, the meta-materialization of tight-binding models is also studied by identifying their equivalent connectivity, enabling us to demonstrate the above exotic NH behavior of the interface experimentally. Our work provides a conceptually rich avenue to construct NH systems for both optics and electronics.
\end{abstract}

\maketitle

\section{\label{sec:level1} Introduction}
Beyond merely being a perturbation, non-Hermitian (NH) contribution to a system can drastically alter its behavior compared to its Hermitian counterpart. One example of this striking phenomenon is the emergence of exceptional points \cite{miri2019exceptional}, at which both eigenvalues and their associated eigenvectors coalesce \cite{Bender_2007,10.1063/1.1418246}. In parallel with those developments, the advent of extending topological phases to NH systems reveals a plethora of unique aspects 
\cite{RevModPhys.93.015005, dai2024non,hu2023non}, such as anomalous bulk-boundary correspondences accompanied by the NH skin effect \cite{PhysRevLett.121.086803,PhysRevLett.121.026808,PhysRevLett.125.126402,WOS:000537039500002,ochkan2024non}. It was widely believed that those phenomena are unique to NH systems. \par
On the other hand, optics and photonics have proven to be the ideal experimental platform to observe the rich physics of NH systems \cite{feng2017non,El-Ganainy2018,RevModPhys.88.035002,RevModPhys.91.015006,feng2025non,rao2024braiding}. This is because that gain (e.g., gain implemented through simulated emission \cite{Rüter2010}) and loss (e.g., material absorption \cite{El-Ganainy:07}) in photonics provide the essential ingredients to construct NH systems in a controllable manner. Under this spirit, recent researches resort to investigating other optical nonconservative processes which can be effectively described by gain/loss, such as the radiation leakage to the surroundings \cite{Zhen2015}, virtual gain \cite{PhysRevLett.124.193901,science.adi1267, kim2025complex}, NH parametric processes \cite{Roy:21}. \par
Can we realize the NH effect in a Hermitian system involving neither gain nor loss? An affirmative answer is obtained in this work, demonstrating the gain/loss or the non-Hermiticity of a system is \textit{not} the necessary condition for NH effects. By definition, the NH terms in the Hamiltonian root in the nonconservative process. So, in principle, the interface between two interacting subsystems is an ideal object for NH physics given the frequent energy and matter flux passing through it. Consequently, it is expected that the NH effect occurs even if the total system comprising subsystems is Hermitian. 

However, the interface should be carefully constructed so that its effective Hamiltonian fulfills the mathematical requirement for NH exotic effects, for example, asymmetric hopping in the NH non-Bloch transport \cite{PhysRevLett.121.086803,PhysRevLett.121.026808,PhysRevLett.125.126402}. Correspondingly, the interface between a topological insulator and a conductor is studied here. This interface has one dimension lower than its constituents, and its effective Hamiltonian consists of: (i) the gapless edge state Hermitian Hamiltonian of the topological insulator; and (ii) the NH self-energy of the conductor that is considered to be a thermal reservoir. As a result, the hopping terms in the effective Hamiltonian are asymmetric due to the unidirectional property of the edge states. 

Here, more explicitly, we study a junction of the Haldane model coupled with a conductor. We theoretically show that this interface Hamiltonian is Hatano-Nelson-like and hopping between lattice sites is asymmetric, leading to the non-Bloch transmission. The experimental realization of this junction poses another challenge since the model is based on the tight-binding approximation. Accordingly, we discuss the electromagnetic meta-materialization of this by analyzing the equivalent site (or node) connectivity between the tight-binding model and the proposed metamaterial, and the results from experiment designed under this scheme are consistent with our prediction of the non-Bloch behavior on the interface. The proposed junction model together with the meta-materialization extends the scope of NH photonics or NH metamaterials.

\section{\label{sec:level1} Non-Bloch transport in the Hermitian system}

\begin {figure*}[ht]
\centering
\includegraphics[width=10cm]{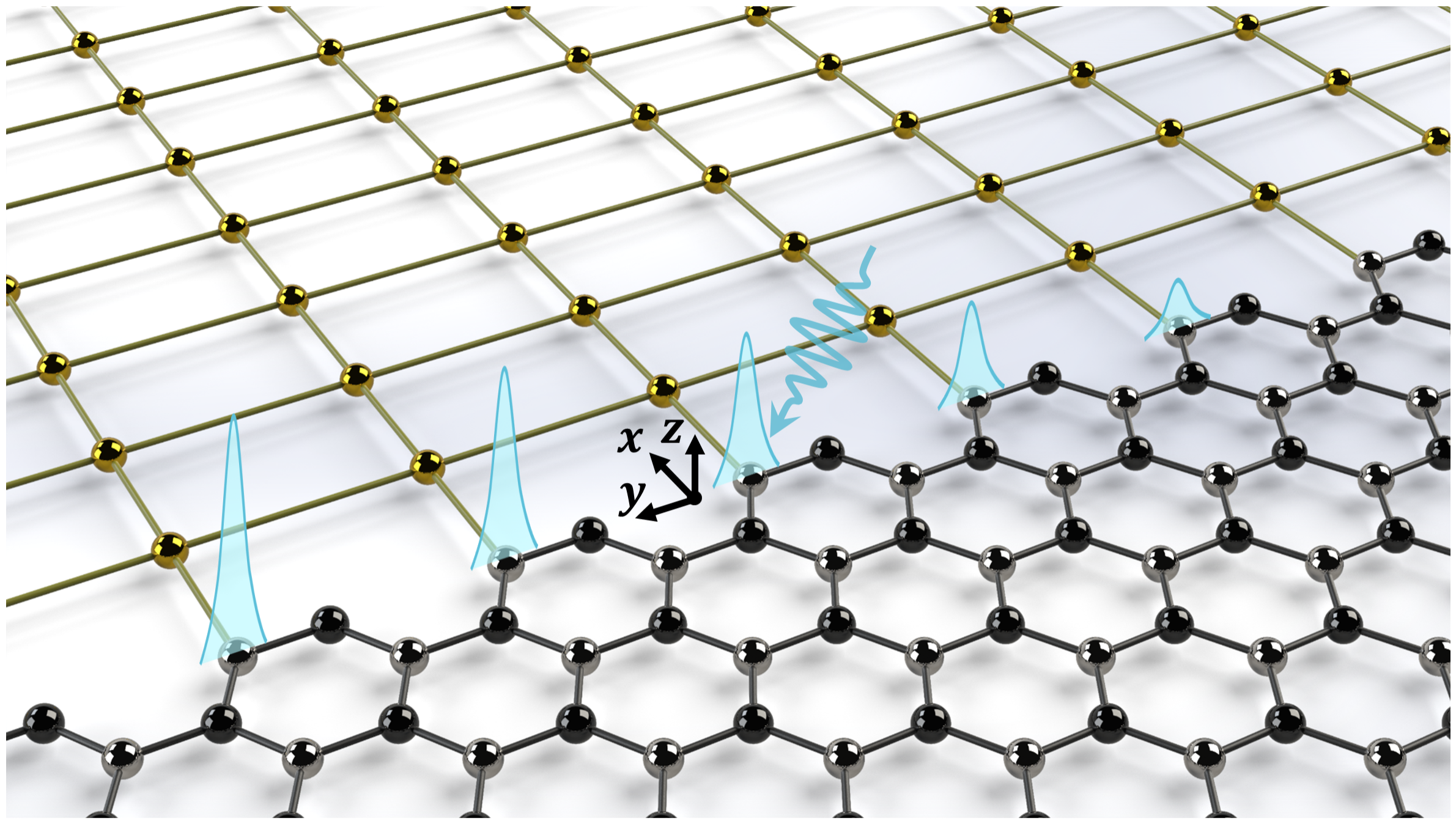}
\caption{Junction between the topological insulator (Haldane model) and the conductor. The next-nearest neighbor hopping $t_2$ is not plotted for simplicity.}
\label{sch}
\end {figure*}

We consider a two-dimensional \added{(2D)} Chern insulator (CI) coupled to a conductor, and the line $x=0$ separates them (see Fig. \ref{sch}\comment{3D plotting}). The whole system is described by the Hamiltonian $\mathcal{H}  = \mathcal{H}_{\mathrm{TI}}+\mathcal{H}_{\mathrm{cond} }+\mathcal{W} $ with hermiticity $\mathcal{H}=\mathcal{H}^\dagger$. 
\begin{equation}
 \mathcal{H}_{\mathrm{TI}} = t_1\sum_{\langle i,j\rangle}c_i^\dagger c_j+it_2\sum_{\langle \langle i,j\rangle\rangle}v_{ij}c_i^\dagger c_j
 \label{H_TI}
\end{equation}
is the Hamiltonian of the Haldane model without breaking the inversion symmetry; $t_{1}$ and $t_2$ are the nearest and next-nearest neighbor hopping of the honeycomb lattice (let's denote the sublattice as A and B); $v_{ij} = \pm1$ depending on the vectors along the two bonds. By referring to the phase diagram of the Haldane model, one can find $\mathcal{H}_{\mathrm{TI}} $ is in the topological phase and the two bands are separated by $\mathbb{G}\equiv (-3\sqrt{3}t_2, 3\sqrt{3}t_2)$, indicating the topological states in the gap. \added{For simplicity, we take the conductor on a square lattice, with its sites near the junction only connecting to the sublattice; the type of lattice or connections, in general, does not affect our conclusion.} The Hamiltonian
\begin{equation}
 \mathcal{H}_{\mathrm{cond}} = \lambda_1 \sum_{\langle i,j\rangle_x} a_i^\dagger a_j + \lambda_2 \sum_{\langle i,j\rangle_y} a_i^\dagger a_j+ \mu  \sum_{i} a_i^\dagger a_i
 \label{H_cond}
\end{equation} 
describes the conductor with its band spanning $\mathbb{B}\equiv [u-2|\lambda_1|-2|\lambda_2|,u+2|\lambda_1|+2|\lambda_2|]$. $\mu $ denotes the onsite energy and $\lambda_{1}$ \added{or $\lambda_{2}$} is the nearest neighbor hopping along $x$ \added{or $y$} of the square lattice. \added{$\mathcal{H}_{\mathrm{cond}}$ is topologically trivial since it has only one band and the total Chern number \deleted{of all bands }must be zero.}
\begin{equation}
\mathcal{W} = \lambda_1\sum_{\langle i,j\rangle_x} a_i^\dagger c_j+h.c.
\end{equation} 
gives the coupling \replaced{between $\mathcal{H}_{\mathrm{TI}}$ and $ \mathcal{H}_{\mathrm{cond}}$}{on the interface}. \par
Several cases are rather trivial or well-studied and hence been excluded here: (i) the excitation energy $\varepsilon$ is in the conduction bands of CI and the conductor, the transport perpendicular to the interface is much like that between conductors; (ii) the excitation energy is in the conduction band of CI and the band gap of the conductor, the perpendicular transport is blocked; (iii) the energy is in their gaps, it is well known that the transport along the interface is unidirectional and the number of the channel equals the Chern number \cite{RevModPhys.82.3045,RevModPhys.83.1057}. \added{Instead,} we consider the transport parallel to the interface when $\varepsilon\in \mathbb{B}\cap \mathbb{G}$. In that case, $\mathcal{H}_{\mathrm{TI}}$ can be reduced to the effective Hamiltonian $H_\mathrm{edge}$ for the edge states, and the reservoir assumption leads to that the effect of  $ \mathcal{H}_{\mathrm{cond}}$ and $\mathcal{W}$ can be replaced by the self-energy $\Sigma $ of the conductors \cite{datta1997electronic,PhysRevResearch.1.012003}. That is
\begin{widetext}
\begin{equation}
\mathcal{H}\to  H_{\mathrm{eff}} =H_{\mathrm{edge} }+\Sigma =
 \frac{t_1}{2} \begin{pmatrix}
  \sin \frac{k_y}{2} &  \frac{1}{2}(1-\mathrm{e}^{ik_y}  ) \\
  \frac{1}{2}(1-\mathrm{e}^{-ik_y}  )& \sin \frac{k_y}{2}
\end{pmatrix}
-\frac{1}{2} \begin{pmatrix}
 \Lambda+ \sqrt{-4\lambda_1^2+\Lambda^2}   & \\
  &0
\end{pmatrix}.
\end{equation}
\end{widetext}
where $\Lambda = \mu -\varepsilon + 2\lambda_2 \cos k_y $. Here, we represent \added{reduced} $H_{\mathrm{eff}}$ in $k_y$-momentum space by noticing the translation symmetry along $y$-axis. \added{In section II of supplementary material (SM), we show that the Haldane model is exactly solvable through Green's function technique, and $H_{\mathrm{edge}}$ is constructed out of the edge state projector.} Note that an extra eigenstate whose eigenvalue is zero appears in constructing $H_{\mathrm{eff} }$, and we need to drop it in the calculation. \added{Also in there, $\Sigma$ is calculated with the help of the transfer matrix.}\par

\begin {figure}[ht]
\centering
\includegraphics[width=\linewidth]{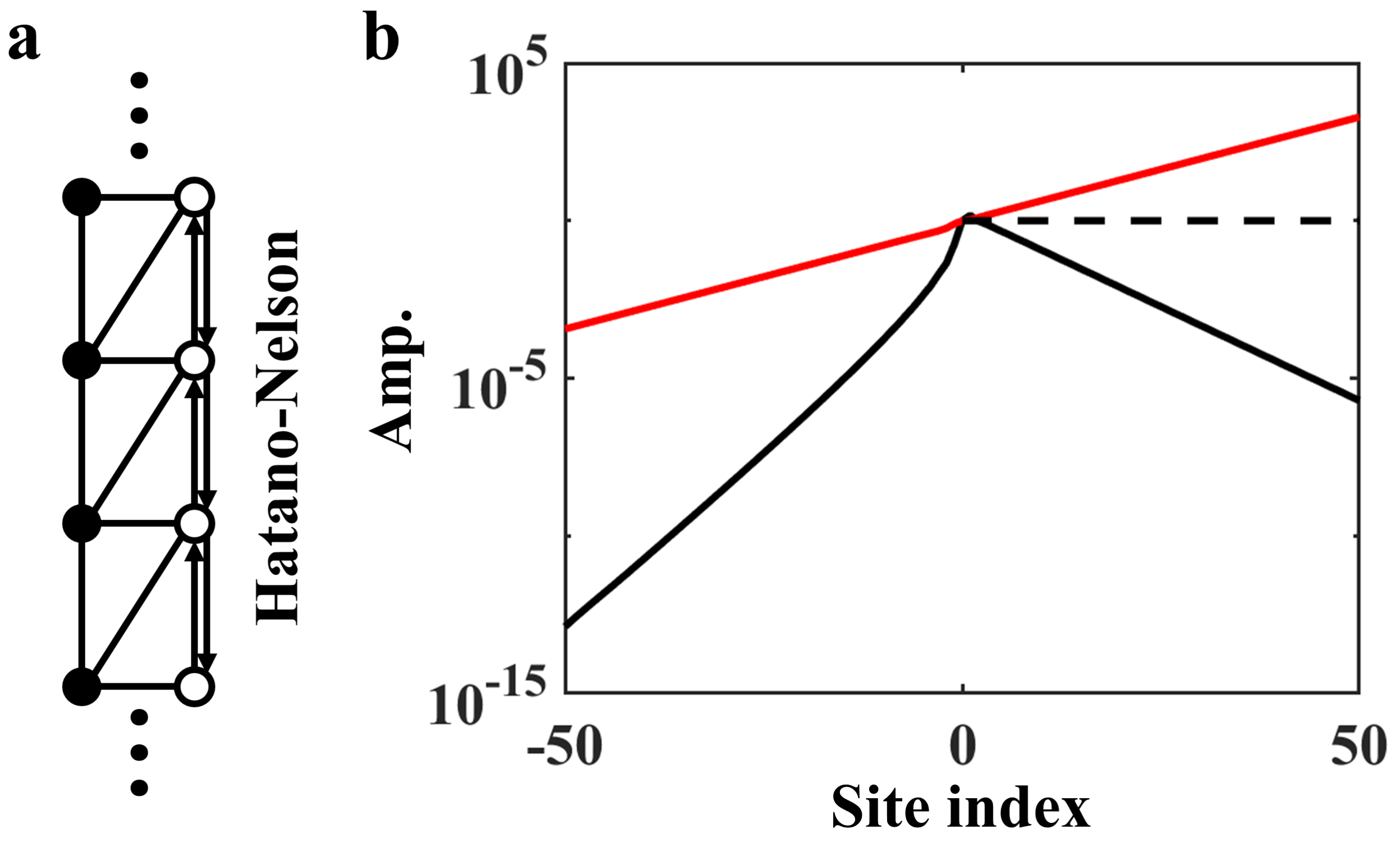}
\caption{Effective model and response profile. (a) Effective chain model at the junction around $k_y=0$. One of the sub-chain is the Hatano-Nelson model. (b) The response profile (from Kwant simulation \cite{Groth_2014}) at the interface when an excitation is injected at the origin. \added{Note that logarithmic scale is on the \replaced{vertical }{y-}axis.} Dashed line: profile of the one-way topological state of the Haldane model. Black solid line: profile at the interface, and the different decay rates \deleted{(left: $0.027$, right: $0.043$)} \added{(left: $1.7$, right: $0.3$)} away from the excitation in the opposite direction suggest the non-Bloch transport. Red solid line: profile when the onsite gain is added on sublattice A.}
\label{effmodel}
\end {figure}

$H_{\mathrm{edge} }=H_{\mathrm{edge} }^\dagger $ but $\Sigma \neq  \Sigma^\dagger 
$ if $\varepsilon\in \mathbb{B}$ hence the effective Hamiltonian $H_{\mathrm{eff} }$ of the interface is NH even though the whole system $\mathcal{H}$ is \textit{Hermitian}. \added{Corresponding to excluded case (iii) mentioned above, $\Sigma =  \Sigma^\dagger $ and $\Sigma\neq 0$ if $\varepsilon \notin \mathbb{B}$, $\Sigma$ will renormalize the edge state, and the dashed line in Fig. \ref{effmodel}b represents the renormalized response profile of the edge state concerning the excitation. Clearly, the wave propagation along the interface is lossless and unidirectional. We remark that the non-Hermiticity lies in the reservoir assumption, which can be roughly understood as that the conductor is significantly larger than the topological insulator, and this assumption fails when considering two comparable subsystems.}\comment{Dashed in Fig. 1c}\par

It is very hard to find a tight-binding chain lattice that corresponds to $H_{\mathrm{eff} }$, so to give it an intuition, we consider the approximation model (e.g., $\sin \frac{k_y}{2}\approx\frac{1}{2} \sin k_y$) around $k_y = 0$ and set \deleted{$t_1=\lambda_1=1$, $\lambda_2=-1.14$ and $\mu=\varepsilon=0$} \added{$\lambda_1=t_1=1$, $\lambda_2=-0.9$ and $\mu=\varepsilon=0$}. Correspondingly, \added{the element in the first row and first column of $H_{\mathrm{eff}}$:}
\begin{equation}
[H_{\mathrm{eff}}]_{11} = 2\xi \sin k_y+2\gamma \cos k_y -1.56i,
\end{equation}
where $2\xi=0.25$ and \deleted{$2\gamma =  1-1.46\mathrm{i}$} \added{$2\gamma =  0.9+1.13i$}. Figure \ref{effmodel}a depicts the approximation tight-binding chain. The \replaced{last term}{complex constant} in the equation \replaced{manifests as}{ results in} a global dissipation\replaced{, while the rest of}{. By inspection,} $[H_{\mathrm{eff}}]_{11}$ is nothing but the Hatano-Nelson model \cite{PhysRevLett.77.570} describing the vortex pinning in superconductors. The hopping in this model is asymmetric, i.e. $t_{ij} = -i\xi+\gamma$ from site j to i while $t_{ji} = i\xi+\gamma$ from site i to j, and $t_{ij}\neq t_{ji}^\ast$; it follows that the bulk states pile up at the boundary, namely, the NH skin effect. To describe the dispersion \added{relation} correctly, one needs to generalize $k_y\in \mathbb{R}$ to the complex domain $k_y\in \mathbb{C}$ \cite{PhysRevLett.121.086803,PhysRevLett.123.066404}, \added{and this can lead to non-bloch transport \cite{PhysRevB.107.064307}}. \par

The response profile $|\varphi|$ at the interface corresponding to an excitation is an indicator of the non-Bloch transport due to $k_y\in \mathbb{C} 
$. \added{In the NH Bloch case where $E\in \mathbb{C} $ and $k_y\in \mathbb{R}$, the profile $\displaystyle |\varphi| \propto \left\{\begin{array}{l} e^{\beta  y},\ y< 0 \\ e^{-\beta  y},\ y>0 \end{array}\right., \beta \in \mathbb{R}$ due to the global dissipation (let the excitation at $y=0$). This means that the wave transport will decay at the same rate away from the excitation. When $E\in \mathbb{R} $ and $k_y\in \mathbb{C}$, the standard non-Bloch case, $|\varphi| \propto e^{\alpha y},\alpha\in \mathbb{R} $. Consequently, the transport is amplified in one direction but is attenuated in the opposite. In general, the superposition\deleted{ed transport} of different \added{transport} types leads to  $\displaystyle |\varphi| \sim \left\{\begin{array}{l} e^{(\alpha+\beta)  y},\ y< 0 \\ e^{(\alpha-\beta)  y},\ y>0 \end{array}\right.$ (more statements see section IV of SM).}\deleted{If the chain eigenenergy $E\in \mathbb{R} $ and $k_y\in \mathbb{C}$, the profile $|\varphi| \propto e^{\alpha y},\alpha\in \mathbb{R} $ because of the skin effect. Inversely, if $E\in \mathbb{C} $ and $k_y\in \mathbb{R}$, 
 $\displaystyle |\varphi| \propto \left\{\begin{array}{l} e^{\beta  y},\ y< 0 \\ e^{-\beta  y},\ y>0 \end{array}\right., \beta \in \mathbb{R}$ due to the global dissipation (let the excitation at $y=0$). Therefore, in general cases $E, k_y\in \mathbb{C}$, 
 $\displaystyle |\varphi| \sim \left\{\begin{array}{l} e^{(\alpha+\beta)  y},\ y< 0 \\ e^{(\alpha-\beta)  y},\ y>0 \end{array}\right.$more statements see Appendix D).} So, the different decay (or growth) rates away from the excitation in the opposite direction indicate the non-Bloch transport. \added{It is worth mentioning that analyzing the response profile in non-Bloch problems is more practical, as it can be directly accessible through regular scattering measurements.}\par

\added{Figure \ref{effmodel}b (black solid line) shows the response profile along the interface from simulation, which corresponds to the general case where $E, k_y\in \mathbb{C}$. Clearly, it is non-Bloch transport even though the entire system is Hermitian. Moreover, in order to compensate for the global dissipation we apply onsite gain to the interface in another simulation, and we have the standard non-Bloch transport (see red line in Fig. \ref{effmodel}b \comment{Dotted line in Fig. 1c}) corresponding to the case where $E\in \mathbb{R} $ and $k_y\in \mathbb{C}$.} \par

The non-Bloch wave at the TI-conductor junction is universal. The reason is two-fold: (i) as part of the junction, the TI does not have to be the type of CI. The localized edge states make it possible to construct an effective one-dimensional system through the interface of two-dimensional materials. So, in principle, the CI (class A) used here can be replaced by other types of TI (e.g. class AII). In a separate work \cite{spearate}, we realize the $\mathbb{Z}_2 $ NH skin effect at the interface between a time-reversal TI and a conductor; (ii) in contrast to Ref. \cite{PhysRevB.107.035306}, no extra requirement for the conductor is adopted here. In there, a relative momentum shift between the reservoir and the nanowire leads to asymmetric hopping. However, this asymmetry in our model results from the interaction between the unidirectional topological states and the reservoir self-energy. 

\section{\label{sec:level1} Experimental demonstration}

\begin {figure*}[ht]
\centering
\includegraphics[width=15cm]{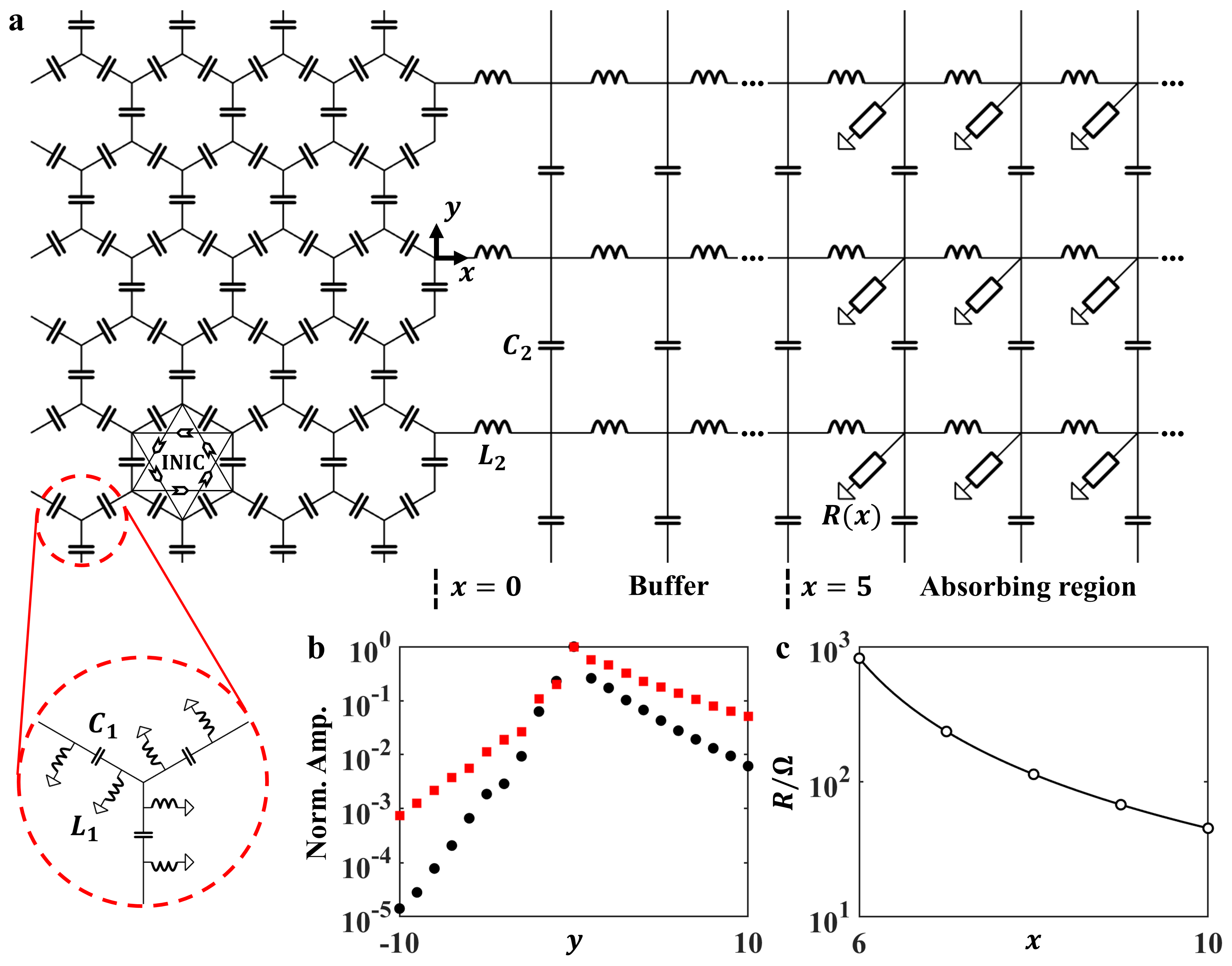}
\caption{Experimental demonstration. (a) Circuit schematics. The INIC is used to realize the next-neighbor hopping (only a few instances are illustrated). The inset shows the grounding configuration\added{, which ensures equal self-admittance} for each node of the honeycomb lattice\deleted{, the nodes surrounding the sample boundary must be grounded carefully to ensure equal self-admittance for each node}. Note that \replaced{one}{the} connecting capacitor $C_1=1 \:\mathrm{nF}$ along with \added{two} grounding inductors $L_1= 3.3 \:\mathrm{\mu H}$ \replaced{between every two adjacent nodes}{at each node} forms a $\Pi$-shaped circuit. $C_2=100 \:\mathrm{pF}$ and $L_2= 3.3 \:\mathrm{\mu H}$. (b) Experimental results of the profile (black dots), indicating the non-Bloch transport. Red squares are the results from CST microwave \added{field-circuit co-}simulation of reverse design. (c) The grounding resistance $R(x)$ varies as $x$. In absorbing region, the complex potentials are realized by grounding all nodes with resistances. }
\label{circuit}
\end {figure*}

\begin {figure}[ht]
\centering
\includegraphics[width=\linewidth]{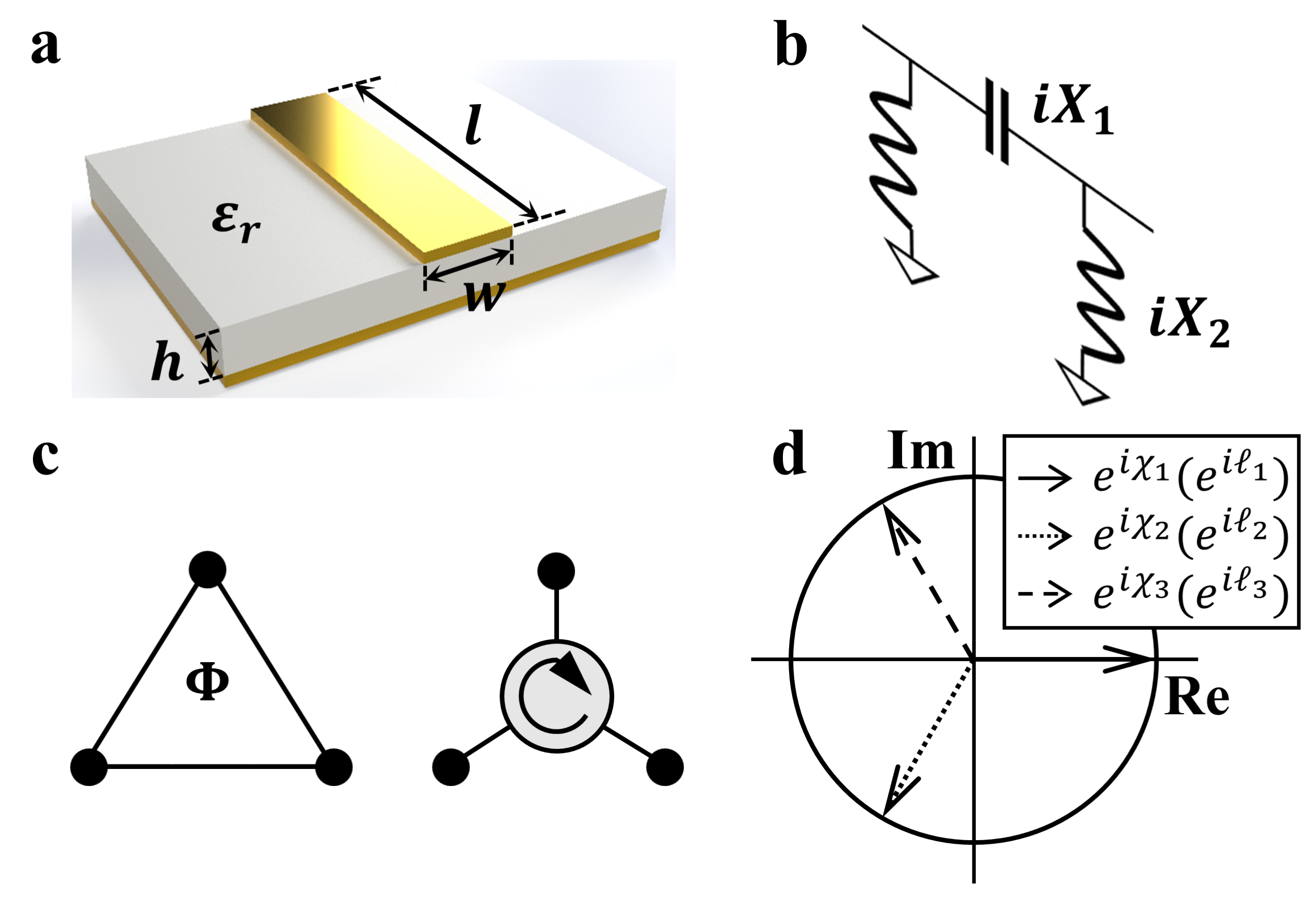}
\caption{Principle of the reverse design and introducing nonreciprocal hopping $t_2$ . (a) Schematic diagram of microstrip waveguide. $w=2.2$ mm, $h=1.2$ mm and $\varepsilon_{\mathrm{r}}=4.6$ lead to the characteristic impedance $Z_0=50\:\Omega$ at 2.4 GHz. (b) $\Pi$-shape equivalent lumped circuit model of the microstrip. (c) Equivalent connectivity between the $\Delta$-configuration of the three-site model and the Y-configuration of a circulator, and their characteristics (d) when $\Phi = 3\pi/2$. $\chi_i\leftrightarrow \ell_i$ \textit{uniquely} specify the hopping relation between the three sites.}
\label{ReverseDesign}
\end {figure}

\textit{Circuit.}---Implementing temporal topolectrical circuit (TTC) \cite{PhysRevB.107.064307,PhysRevLett.132.016605} theory, we can construct a circuit lattice with those correspondences, Hamiltonian $\mathcal{H}$ $\leftrightarrow $ admittance $Y$ of circuits, eigenvalues of $\mathcal{H}$ $\leftrightarrow $ self-admittance $Y_0$ of circuits,\added{ and eigenmodes of $\mathcal{H}$ $\leftrightarrow $ voltage on the circuit nodes}. \deleted{The spirit of reverse design is that a waveguide with length (e.g., a microstrip line in Fig. \ref{ReverseDesign}a) is equivalent to a lumped circuit model (Fig. \ref{ReverseDesign}b). This equivalence enables one to `update' a circuit lattice to a metamaterial in wave systems very effectively. More details about design including the graphene circuit, the converted waveguide metamaterial and its optimization can be found in Appendix F.} \added{Figure \ref{circuit}a illustrates the schematic of the designed circuit lattice and Fig. \ref{circuit}b (black dots) shows the voltage response profile under a current excitation from the experiment. Without the square lattice, which acts as the conductor $\mathcal{H}_{\mathrm{cond} }$, the voltage propagates in a unidirectional manner. However, when it is introduced, the voltage wave propagates in both directions but with different decay rates, indicating the non-Bloch transport. In the experiment, the time-reversal-breaking term $t_2$ is realized by the negative impedance converter with current inversion (INIC, Fig. \ref{circuit}a). }
\par

In the model of $\mathcal{H}$, we assume that the size of the conductor $\mathcal{H}_{\mathrm{cond} }$ is semi-infinite such that it serves as a reservoir. It is very cumbersome to prepare the circuit sample in that size. Instead, one can reduce the semi-infinite geometry to a finite lattice with proper boundary condition, i.e., the perfect absorbing boundary condition using power-growth complex onsite potentials that completely absorb the incident wave. That gradient complex potentials are realized by grounding the nodes with resistances (see Fig. \ref{circuit}c). More circuit experiments refer to section VIII of SM. We remark that the boundary complex potential is not the origin of non-Bloch transport and the semi-infinite argument may also be mitigated by a fairly large lattice.

\textit{Electromagnetic metamaterials.}---\replaced{The normal design of topological metamaterials costs substantial computational resources: start with geometry and material parameters; then calculate the band structure by finite element simulation; if the band is not topological, tune the parameter and start over. On the contrary, the proposed reverse design is non-iterative and in principle requires minimal computational resources: it starts with a tight-binding model, then designs a circuit network based on the TTC, and finally generalizes the network to the electromagnetic/optical metamaterials.}{For the normal design  procedure of topological metamaterials, one starts with some necessary parameters, such as lattice parameters, material parameters, unit cell structure, and so on; with those parameters, the dispersion relation of the metamaterials can be calculated  with the help of the finite element simulation; if the band from the simulation is not topological, we tune the parameters and start over. That iterative  procedure costs a huge of computing resources because of lack of generality, and the design may pose a challenge to sample fabrication (e.g., \cite{Lu2013}). The starting point of the reverse design proposed here is the tight-binding model, and in principle, the design does not need the aid of finite element simulation.} As a proof of concept, we illustrate this idea by realizing the junction mentioned above in the S band ($2-4$ GHz) range, and the design scheme is readily generalized to the optical range.\par

\added{It takes two steps to generalize the lumped circuit network mentioned above to electromagnetic/optical metamaterials. This is because INIC that breaks time-reversal symmetry does not have a counterpart in that frequency range, step two will handle that. We remark that compared to the existing topolectrical circuits \cite{Lee2018,Imhof2018,WOS:000882542700001}, TTC is more suitable to be the starting point of the reverse design, because the impedance measurements required by existing circuits are out of reach in scattering experiments while scattering measurements are more accessible for electromagnetic metamaterials.}\par

\added{First, consider the circuit network without INIC. The key observation is that each $\Pi$-shape circuit in the network is equivalent to a segment of waveguide with electrical length $kl$, where $k$ is wavenumber in the waveguide and $l$ is the physical length. The admittances of the grounding inductors and the connecting capacitor are 
\begin{equation}
iX_1 = -i\frac{1}{Z_0\sin kl},\ iX_2 = i\frac{1}{Z_0}\tan \frac{kl}{2},
\end{equation}
respectively\comment{double check}, where $Z_0$ is the characteristic impedance of the waveguide. One can easily check this equivalence by calculating the two-port scattering matrix of the $\Pi$-shape circuit (Fig. \ref{ReverseDesign}b) and the waveguide \cite{10.5555/27077}. Here, we choose the waveguide as the microstrip line (Fig. \ref{ReverseDesign}a). 
In the honeycomb lattice, we set $kl=3\pi/2$, which corresponds to $C= 1.3 \:\mathrm{pF}$ and $L= 3.3 \:\mathrm{nH}$ at 2.4 GHz\comment{needs to be filled}. In the square lattice we set $kl=3\pi/2$ for $\lambda_1$ and $kl=9\pi/2$ for $\lambda_2$, which also corresponds to $C= 1.3 \:\mathrm{pF}$ and $L= 3.3 \:\mathrm{nH}$ at 2.4 GHz, with grounding and connecting components exchanged\comment{needs to be filled}. 
Note that the values of lumped elements should be scaled up/down owing to the frequency change.} \par
Next, we show how to introduce the next neighbor T-breaking hopping $t_2$ of $ \mathcal{H}_{\mathrm{TI}}$ which is critical to the Haldane model. It is well-known this T-breaking term can be realized by nonreciprocal devices \cite{Wang2009,PhysRevLett.114.114301,PhysRevLett.122.247702,Zhang2021}, e.g., a RF circulator. In reality, a nonreciprocal device has very limited bandwidth, for example, the typical bandwidth is a few hundreds MHz for a RF circulator working at GHz, thus the frequency range of resonance-based metamaterials emulating the tight-binding Haldane model should be in this bandwidth. This frequency concern should not bother us since our TTC or TTC-based design operates at a fixed frequency and is not resonance-based. According to our previous works \cite{PhysRevB.107.064307,PhysRevLett.132.016605}, it looks like introducing the hopping $t_2$ can be easily achieved by finding out the admittance of the RF circulator, however, in general, the circulator does not have a well-defined admittance (Part A in section V of SM).\par

\begin {figure}[ht]
\centering
\includegraphics[width=\linewidth]{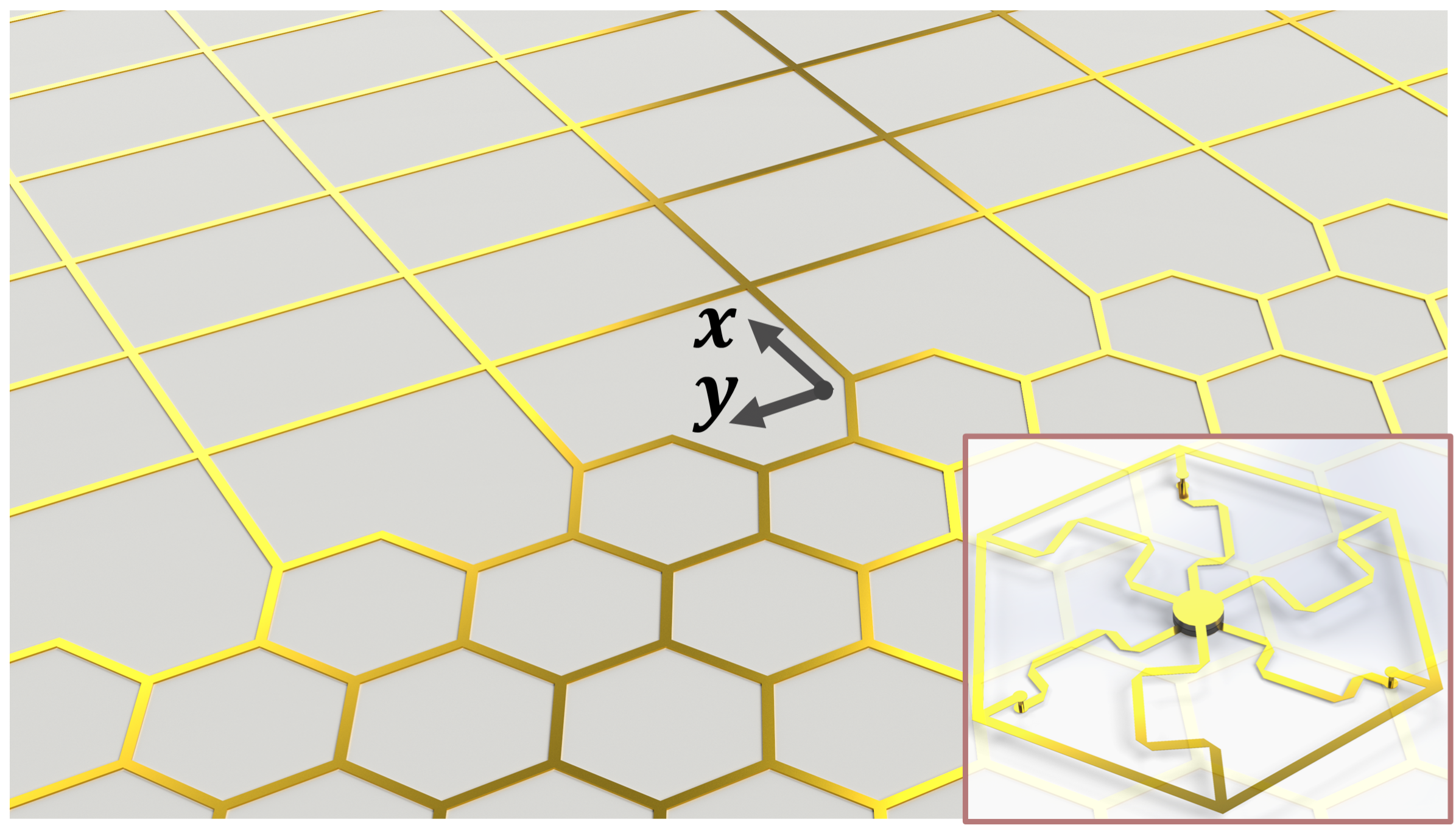}
\caption{Metamaterial realization of non-Bloch transport under reverse design. Metamaterial junction consists of microstrip lines. \deleted{The inset, where substrates are not plotted for clarity, shows the non-reciprocal coupling $t_2$ using the three-port circulator.}\added{For clarity, three-port circulators and their connecting lines that lead to the non-reciprocal coupling $t_2$ are only shown in the inset, where substrates and the middle layer of ground plane are not plotted.} The perfect match layer (PML) boundary condition is applied to mimic the semi-infinite conductor.}
\label{Metamaterial}
\end {figure}

To solve that, let's first understand the hopping in a geometric manner. Without loss of generality, we consider a minimal model containing three identical sites with hopping $t_2$ connecting them. The flux $\Phi$ in the plaquette leads to a phase $\theta=\frac{\Phi}{3}$ in $t_2$, and the Hamiltonian is
\begin{equation}
\mathcal{H}_\Delta /t_2 = (\Lambda_1+  \Lambda_4+\Lambda_6)\cos \theta+ (\Lambda_2-  \Lambda_5+\Lambda_7) \sin\theta
\end{equation}
where $\Lambda_i$ are the Gell-Mann matrices. When $\Phi=0$, $\mathcal{H}_\Delta$ respects D3h symmetry, thus it has two-fold degeneracy eigenvalues. When pumping the flux, the symmetry is lifted down to C3, resulting in the splitting of the degeneracy (note that when $\Phi = n\pi$ with $n\in \mathbb{Z} $, $\mathcal{H}_\Delta$ respects D3h up to a gauge). So, we can use the frame of $\mathrm{u}=(\mathbf{v}_1,\mathbf{v}_2,\mathbf{v}_3 ) $, $\mathbf{v}_1 = (1,1,1)^{\mathrm{T} },\mathbf{v}_2 = (1,\alpha,\alpha^2)^{\mathrm{T} },\mathbf{v}_3 = (1,\alpha^2,\alpha)^{\mathrm{T} }$ with $\alpha = e^{\frac{2\pi i}{3} }$. Performing a bilinear transform, we have 
\begin{equation}
\mathcal{S}=\frac{ i +\mathcal{H}_\Delta }{i-\mathcal{H}_\Delta  }.
\end{equation}
$\mathcal{S}$ is unitary: $\mathcal{S}\mathcal{S}^{\dagger }=1$, and in the frame of $\mathrm{u}$: $\mathrm{u}^{-1}\mathcal{S}  \mathrm{u} =\mathrm{diag}(e^{i\chi _1},e^{i\chi _2},e^{i\chi _3} ) $. So, characteristics $\chi _{1,2,3}$ \textit{uniquely} specify the hopping relation between the sites (see Fig. \ref{ReverseDesign}d). On the other hand, compared to the admittance, the scattering-parameter of a circulator is always well-defined. In section V of SM, we demonstrate that three similar characteristics $ \left ( \ell _1,\ell _2,\ell _3\right ) $ in a circulator uniquely determine the connectivity of the three sites; this gives us an correspondence of T-breaking hopping between tight-binding models and the metamaterials (Fig. \ref{ReverseDesign}c): $\chi_i\leftrightarrow \ell_i$. In section VI of SM, the simulation establishes that the topological metamaterial designed under this correspondence supports the unidirectional edge state just as the Haldane model does.\par 
\deleted{Last, in the model of $\mathcal{H}$, we assume that the size of the conductor $\mathcal{H}_{\mathrm{Cond} }$ is semi-infinite such that it serves as a reservoir. The metamaterial construction of $\mathcal{H}_{\mathrm{Cond} }$ is similar to that of $ \mathcal{H}_{\mathrm{TI}}$ when $t_2=0$, but it is very cumbersome to prepare the sample in that size. Instead, one can reduce the semi-infinite geometry to a finite lattice with proper boundary condition, i.e., the perfect absorbing boundary condition using power-growth complex onsite potentials that completely absorb the incident wave (see Appendix G). The complex potentials are realized by grounding the nodes with resistances (see Fig. \ref{Metamaterial}b). We remark that the boundary complex potential is not the origin of non-Bloch transport and the semi-infinite argument may also be mitigated by a fairly large lattice.}\par
Figure \ref{Metamaterial} shows the finial RF design of the junction with microstrip lines, and Fig. \ref{circuit}b (red squares) gives the electric field profile with respect to an excitation (from CST \added{field-circuit co-}simulation). Clearly, the metamaterial junction exhibits the property of the non-Bloch transport, which is consistent with the prediction.

\section{\label{sec:level1} Outlooks}

In summary, we provide an alternate approach to realize non-Hermitian systems in electromagnetics in a gain\added{/loss}-free fashion. The reverse design procedure, i.e., the tight-binding models$\rightarrow $temporal topolectrical circuits$\rightarrow $electromagnetic metamaterials, offers a prevalent paradigm to design the topological band of the metamaterials.

\begin{acknowledgments}

We gratefully acknowledge the insights of and discussions
with Fuxin Guan, Bo Zhen, Ce Shang, Qinghua Song and Li Ge.\par
This work was supported by National Key Research \& Development Program of China 
(Grant No. 2023YFB3811400) and National Natural Science Foundation of China (Grant No. 52273296).

\end{acknowledgments}

\bibliography{ref.bib}

\foreach \x in {1,...,16}
{%
\clearpage
\includepdf[pages={\x,{}}]{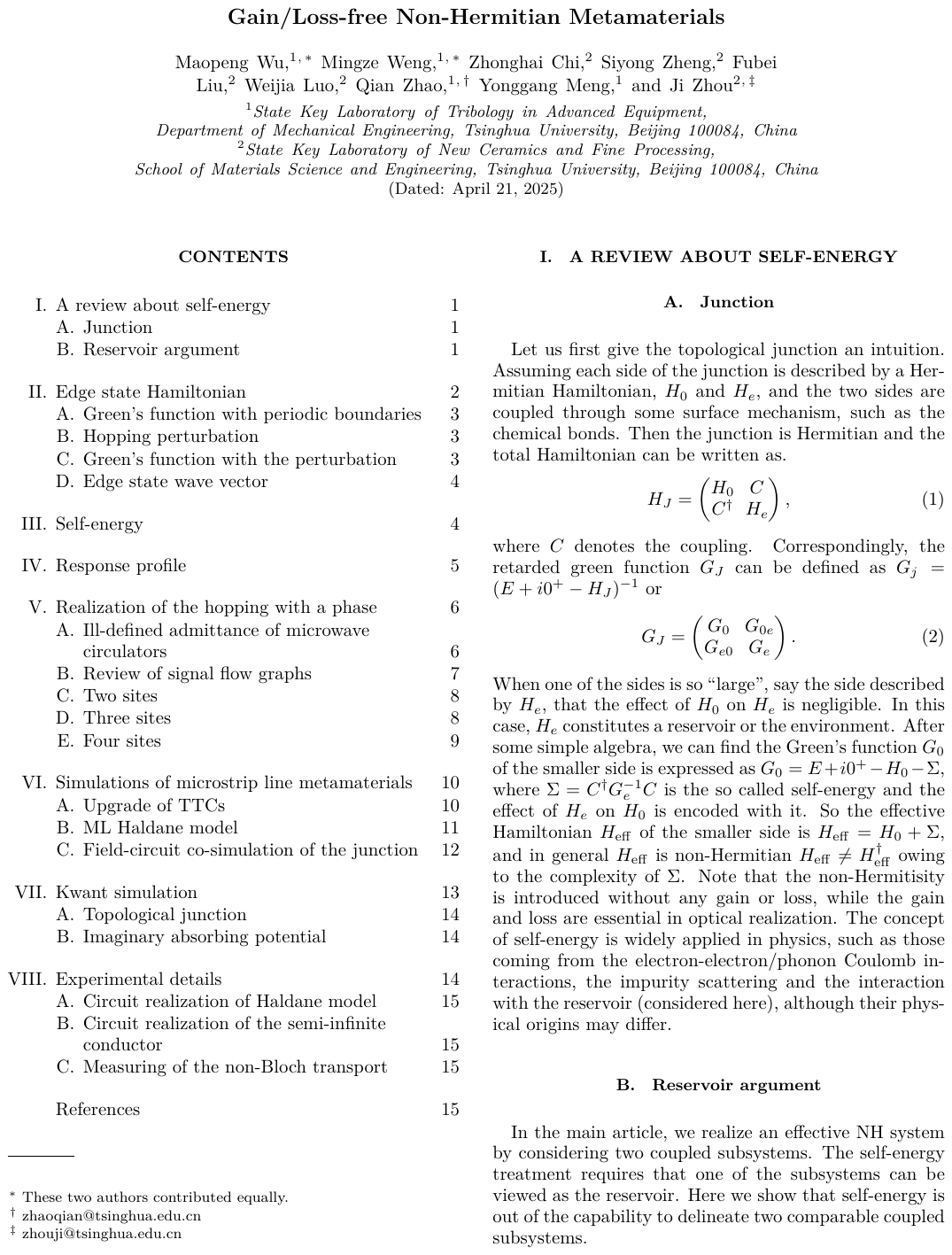}
}

\end{document}